LEPTON SELF-MASS, HIGGS SCALAR AND HEAVY QUARK MASSES[1]

R. DECKER AND J. PESTIEAU

Institut de Physique Théorique
Université de Louvain,
BELGIUM

Assuming concellation of lepton self-mass ultraviolet divergences in the SU(2) × U(1) gauge theory, we predict a heavy quark $t$ with mass, $m_t = 85$ Gev (100 Gev), and a Higgs scalar $\psi$ with mass, $M_\psi = 90$ Gev (140 Gev).

Postal address : Chemin du Cyclotron 2, B-1348 Louvain-la-Neuve, Belgium.

---

[1] Presented at the DESY Workshop, October 22-24, 1979.

In the last two years, it has been seen that experimental data [1] agreed very well with the Weinberg-Salam (W-S) [2] model, the SU(2) × U(1) gauge theory of weak and electromagnetic interactions.

Fundamental fermions (point-like spin 1/2 leptons and quarks) are assumed to interact as left-handed weak isodoublets while the charged leptons and quarks are interacting as right-handed weak isosinglets in successive generations :

$$(\nu_e\ e)_L, (\nu_\mu\ \mu)_L, (\nu_\tau\ \tau)_L \ldots\ ; e_R, \mu_R, \tau_R \ldots\ .$$

$$(u\ d')_L, (c\ s')_L, (t\ b')_L \ldots\ ; u_r, c_r, t_r, \ldots\ , d'_r, s'_r, b'_r \ldots\ .$$

Neutrinos have no right-handed weak isosinglets and are therefore massless. The primes indicate the Cabbibo type intergenerations mixing [3] of quark types. Electromagnetic interactions are mediated by the photon $\gamma$ and weak interactions are assumed to be mediated by three massive intermediate vector bosons $W^\pm$, $Z^\circ$. In order that the theory be renormalizable, [4] fermions and intermediate vector bosons are believed to be massive via the Higgs mechanism [5].

By measuring the electric charge $e$, in Coulomb scattering, the Fermi coupling constant, $G_F$, in $\mu$-decay and the ratio of the elastic $\nu_\mu e$ and $\overline{\nu_\mu} e$ cross sections, it is possible to determine the gauge coupling $g$, the charged intermediate vector boson mass, $M_w$, and the Weinberg angle $\theta_W$.

Furthermore comparison between charged and neutral current experiments allows to measure the parameter $\rho = \dfrac{M_w^2}{M_z^2 \cos^2 \theta_W}.$

The result is $\rho = 1.004 \pm 0.018$ [6] close to 1 $\rho = 1$ is obtained in the W-S model because the SU(2) × U(1) symmetry is broken by an SU(2) doublet of Higgs scalars [7].

The model has other free parameters [3]; the fundamental fermion masses, $m_j$, the Higgs mass, $M_\psi$, Cabbibo like angles and CP-violation phases. Because the model is renormalisable [4], radiative corrections can be computed and this allows, in some cases, to put limits on heavy fundamental fermion masses [8] and on Higgs boson mass [9].

Interactions between fermions, gauge bosons and Higgs boson fields develop corrections to the fermion masses, functions of the mass of the corresponding fields but this radiative corrections generally cannot be calculated because of ultraviolet divergencies [10].

Forty years, ago, Stueckelberg [11] conjectured that a classical electron could have a finite selfenergie if, in addition to the electromagnetic interactions, there is a new type of interaction with the <u>same</u> strength as the electromagnetic one. In the present paper, we apply this idea to the W-S model with one SU(2) doublet of Higgs scalars.

Previous attempts have been done to find if the Stueckelberg's conjecture could be implemented in the frame of Q.E.D. and weak intermediate vector boson models. In view of cancellation of logarithmic divergences of electromagnetic self mass, were only considered logarithmic divergences of the weak selfmasses. Taking in account just charged weak currents, Terazawa [12] was the first to notice that due to the presence of $\gamma_5$ matrix in the lepton weak boson coupling it was possible to have a negative sign between the logarithmic weak and electromagnetic divergencies. In Ref. [13] the same idea has been applied to the

Glashow model [14] and the logarithmic divergence cancellation occures if

$$\sin^2 \theta_W = 0.214 \tag{1}$$

in good agreement with present data.

Let us now look to the W-S model. All diagrams in Fig. (1) contribute to the lepton selfmass at the lowest order. One can easily check that the selfmass is independent of gauge as expected for an observable quantity. If we perform the computation in $\xi$ gauges, [15] quadratic divergent part of fermion selfmasses is developped by tadpoles only (Fig. (1b)). Cancellation may be possible because fermion loops have opposite sign to boson loops. All diagrams in Fig. (1a) and Fig. (1b) contribute to the logarithmic divergencies and again cancellation in possible.

We impose that radiative corrections to leptons should be finite (only leptons and <u>not</u> quarks because up to now, they are the only free fundamental particles observed in nature). First of all, we note that neutrinos are massless at any order in the W-S model. For charged leptons we are going to show that the divergence cancellation will occur if there is at least one fermion heavier than the charged weak boson $W^+$. Furthermore, if there is one quark heavier than $W^+$, we will show that (1) it is the only one, to be identified with $t$, the partner of the $b$ quark, and (2) all leptons are much lighter than $W^+$. The $t$ quark is then predicted, with $\sin^2 \theta_W = 0.22$ to be either $m_t = 85$ Gev or $m_t = 100$ Gev and correspondingly, the Higgs boson mass to be either $M_\psi = 90$ Gev or $M_\psi = 140$ Gev.

We now derive our results. Charged lepton $\ell$ selfmass, has contributions : (1) $\delta m_\ell^{(a)}$

from its self interactions with the photon $\gamma$ the weak bosons, $W^+$, $W^-$, $Z^-$, the Higgs scalar $\psi$ and the unphysical scalars, $s^\pm, \chi$, (Figs (1a)) and (2) $\delta m_\ell^{(b)}$ from tadpole – loops of $W^+, W^-, Z^\circ, s^+, s^-, \psi, \chi$, leptons, coloured quarks and ghosts $\phi^+, \phi^-$ and $\phi^Z$ (Fig. (1b)).

The lowest order lepton selfmass can now be computed [16] and their divergencies are given by

$$\Delta m_\ell = \Delta m_\ell^{(a)} + \Delta m_\ell^{(b)} \qquad (2)$$

for $\ell \neq \nu$, with

$$\Delta m_\ell^{(a)} = \frac{e^2}{64\pi^2} \left\{ 6\sec^2\theta_W - \frac{1}{2}\xi_z^{-1}\sec^2\theta_W \text{cosec}^2\theta_W \right.$$
$$\left. -\xi_W^{-1}\text{consec}^2\theta_W + O\left(\frac{m^2 e}{M_W^2}\right) \right\} m_\ell \, \ell n\lambda^2 \qquad (3)$$

and

$$\Delta m_\ell^{(b)} = \frac{e^2}{64\pi^2} \left\{ -\left[3(2+\sec^2\theta_W)x^{-1} + 3 - 4x^{-1}y\right] \cdot \right.$$
$$\text{cosec}^2\theta_W \, m_\ell \frac{\lambda^2}{M_W^2} + \left[\frac{3}{2}x - 4x^{-1}z + \right.$$
$$\left. 3(2+\sec^4\theta_W)x^{-1} + \frac{1}{2}\xi_z^{-1}\sec^2\theta_W + \xi_W^{-1}\right] \text{cosec}^2\theta_W \, m_\ell \, \ell n\lambda^2 \qquad (4)$$

$\lambda$ is the common cutoff and we defined

$$x = \frac{M_\psi^2}{M_W^2} \; ; \; y = \sum_j \left(\frac{m_j}{M_W}\right)^2 \; ; \; z = \sum_j \left(\frac{m_j}{M_W}\right)^4 \qquad (5)$$

where $j$ indicates the summation on all charged leptons and all quark flavours and colours, $\xi_W$ and $\xi_Z$ are the gauge parameters [15] associated with $W^\pm$ and $Z^\circ$ respectively.

Radiative corrections to lepton masses will be finite if the coefficients multiplying $\lambda^2$ and $\ell n\lambda^2$ in Eq. (2) are zero (i.e. if $\Delta m_\ell = 0$)

$$\lambda^2 : \; x + 2 + \sec^2\theta_W - \frac{4}{3}y = 0 \qquad (6)$$

$$\ell n \lambda^2 : \frac{1}{2}x^2 + 2x\mathrm{tg}^2\theta_W + 2 + \sec^4\theta_W - \frac{4}{3}z = 0 \tag{7}$$

where we have neglected terms of order $(m_e/M_W)^2$. With the experimental value of $\sin^2\theta_W$, we note [17] that $z > y$ subtracting Eq. (6) from Eq. (7). From the definition of $y$ and $z$, we deduce that there exists at least one fundamental fermion with a higher mass than the one of $W^{\pm}$. Assuming this fermion to be quark [18], $y$ and $z$ may be parametrized in the following way. Define $t = \left(\frac{m_q}{M_W}\right)^2$, with $q$, the quark heavier than $M_W(t > 1)$. Then

$$y \equiv 3(t + \epsilon) \tag{8}$$

$$z \equiv 3(t^2 + \epsilon') \tag{9}$$

with the factor 3 from colour,

$$\epsilon \equiv \frac{1}{3}\sum_{j \neq q}\left(\frac{m_j}{M_W}\right)^2 \tag{10}$$

$$\epsilon' \equiv \frac{1}{3}\sum_{j \neq q}\left(\frac{m_j}{M_W}\right)^4 \tag{11}$$

Eliminating $x$ between Eqs (6) and (7), one gets

$$t^2 - t(4 - \sec^2\theta_W - 4\epsilon) + 2\epsilon^2 - \epsilon' - \epsilon(4 - \sec^2\theta_W) + 2 - \frac{1}{8}\sec^4\theta_W = 0 \tag{12}$$

By definition of $\epsilon$ and $\epsilon'$, we have $\epsilon^2 > \epsilon'$ and from Eq. (12) : we can write the inequality.

$$t^2 - t(4 - \sec^2\theta_W - 4\epsilon) + \epsilon^2 - \epsilon(4 - \sec^2\theta_W) + 2 - \frac{1}{8}\sec^4\theta_W < 0 \tag{13}$$

This inequality implies that its discriminant $\Delta$ is positive and that $t$ is between the two roots :

$$\Delta^2 \equiv 12\epsilon^2 - 4(4 - \sec^2\theta_W)\epsilon + 8 + \frac{3}{2}\sec^4\theta_W - 8\sec^2\theta_W > 0 \tag{14}$$

if

$$\epsilon < \frac{1}{6}(4 - \sec^2 \theta_W - \delta) \qquad (15)$$

$$\epsilon > \frac{1}{6}(4 - \sec^2 \theta_W + \delta) \qquad (16)$$

with

$$\delta^2 = 16 \sec^2 \theta_W - 8 - \frac{7}{2} \sec^4 \theta_W > 0 \qquad (17)$$

Then

$$\frac{1}{2}(4 - \sec^2 \theta_W - 4\epsilon - \Delta) < t < \frac{1}{2}(4 - \sec^2 \theta_W - 4\epsilon + \Delta) \qquad (18)$$

Eqs (16) and (18) together imply [19] that $t < 1$ in contradiction with what has been assumed. Then the only solutions of $\epsilon$ are given by Eq. (15). Because of $\epsilon \geq 0$, we have to assume $\sin^2 \theta_W < 0.25$ to fulfill Eq. (15). Experimentally $\sin^2 \theta_W > 0.20$. This means that $\epsilon < 0.032$. But by definition, $\epsilon < 1/3$ indicates that there is no other fermion except $q$ heavier than the weak boson $W^+$. Experimentally [20] it seems to be no $t$ quark below 15 Gev. So if there is a quark with a mass between 15 Gev and $M_W$, $\epsilon > 0.036$. To have a fourth quark generation, we therefore need $\epsilon > 0.072$ in contradiction with our result $\epsilon < 0.032$.

Thus we must identify $q$, the heavy quark with the $b$ partner. $t$, the top quark.

With all known quarks $\epsilon \approx 0.004$ and we may neglect $\epsilon^2$, $\epsilon'$ in Eq. (12) and we get

$$t = 2 - \frac{1}{2}\sec^2\theta_W - 2\epsilon \pm \frac{1}{2}\left[8(1 - \sec^2\theta_W) + \frac{3}{2}\sec^4\theta_W - 4\epsilon(4 - \sec^2\theta_W)\right]^{1/2} \qquad (19)$$

$$x = 6 - 3\sec^2\theta_W - 4\epsilon \pm \left[8(1 - \sec^2\theta_W) + \frac{3}{2}\sec^4\theta_W - 4\epsilon(4 - \sec^2\theta_W)\right]^{1/2} \qquad (20)$$

Choosing $\sin^2 \theta_W = 0.22$ we obtain

$$m_t = 85 \text{ Gev} \quad \text{and} \quad M_\psi = 90 \text{ Gev},$$

or

$$m_t = 100 \text{ Gev} \quad \text{and} \quad M_\psi = 140 \text{ Gev}.$$

It is interesting to note that the result of Ref. [13] given in Eq. (1), is obtained if we impose, in Eq. (2) $\Delta m_\ell^{(a)} = 0$ and $\Delta m_\ell^{(b)} = 0$ separately in the 't Hooft gauge (i.e. $\xi_W = 1$ and $\xi_Z = 1$). This is compatible with the assumption made in this paper, i.e. $\Delta m_\ell = 0$ (which is independent of the $\xi$ gauge).

**REFERENCES**

1) Proceedings of *the XIXth International Conference on High Energy Physics*, Tokyo (August 1978). Proceedings of *the International Conference on Neutrinos, Weak Interactions and Cosmology*, Bergen (June 1979).

2) S. Weinberg, Phys. Rev. Lett. **19**, 1264 (1967); **27**, 1688 (1971); A. Salam, in *Elementary Particle* Physics, ed. by N. Svartholm (Almqvist and Wikselis, Stockholm, 1968), p. 367; S.L. Glashow, J. Iliopoulos and L. Maiani, Phys. Rev. **D2**, 1285 (1970).

3) See for example : H. Harari, Physics Reports **42C**, 235 (1978).

4) G. 't Hooft, Nucl. Phys. **B35**, 167 (1971). For reviews and further references, see M. Veltman, in *Proceedings of the 6th Internatiional Symposium on Electron and Photon Interactions at High Energies*, ed. by H. Rollnik and W. Pfeil (North-Holland), American Elsevier (1974), p. 429.

5) F. Englert and R. Brout, Phys. Rev. Lett. **13**, 321 (1964), P.W. Higgs, Phys. Lett. **12**, 132 (1964); Phys. Rev. Lett. **13**, 508 (1964).

6) See for example P. Langacker, Bergen (1979) in Ref. (1).

7) The relation $M_W^2 = M_Z^2 \cos^2 \theta_W$ has been obtained as a consequence of the Glashow model (S.L. Glashow, Nucl. Phys. **22**, 579 (1961)) by J. Pestieau and P. Roy, Phys. Rev. Lett. **23**, 349 (1969), Eq. (5).
Note this relation is <u>not</u> typical for a gauge theory with spontaneous breakdown through elementary Higgs scalars. In this reference : $\theta = \theta_W + \frac{\pi}{2}$.

8) M. Veltman, Cargèse Lectures, July 1979; Nucl. Phys. **B123**, 89 (1977); M. Chanowitz, M. Furman and I. Hinchliffe Phys. Lett. **78B**, 285 (1978); Nucl. Phys. **B153**, 402 (1979).

9) S. Weinberg, Phys. Rev. Lett. **36**, 294 (1976); A.D. Linde, Pis'ma Zh. Eksp. Teor. Fiz. **23**, 73 (1976) (Translation in Jept. Lett. **23**, 64 (1976)); Pham Quarg Hung, Phys. Rev. Lett. **42**, 873 (1979); H.D. Politzer and S. Wolfram, Phys. Lett. **82B**, 242 (1979); **83B**, 421 (1979) (E).

10) Of course mass renormalization makes ther integrals finite, but the physical mass remains a free parameter of the theory and so the radiative corrections are not calculable as a function of the bare mass.
See, however, S. Weinberg, Phys. Rev. **D7**, 2887 (1973).

11) E.C.G. Stueckelberg, Nature **144**, 118 (1939). See also, S. Sakata, O. Hara, Prog. Theor. Phys. **2**, 30 (1947).

12) H. Terazawa, Phys. Rev. Lett. **22**, 254 (1969); **22**, 442(E) (1969); Phys. Rev. **D1**, 2951 (1970).

13) J. Pestieau and P. Roy, see Ref. (7).

14) S.L. Glashow, see Ref. (7).

15) See for example, K. Fujikawa, B.W. Lee and A.I. Sanda, Phys. Rev. **D6**, 2963 (1972) or S. Weinberg in Ref. (10).

16) See, for example, S. Weinberg in Ref. (10).

17) Eq. (7) is satisfied for any value of $x(\geq 0)$ when $\sin^2\theta_W \geq 0.15$.

18) If we assume this fermion, heavier than $W^+$ to be a heavy charged lepton $L$, we can show that its mass will be at least twice the vector boson mass.

19) Eq. (18) is satisfied when $0.75 \geq \sin^2\theta_W \geq 0$.

20) D.P. Barber *et al.*, Phys. Lett. **85B**, 463 (1979).

Figure Caption.

Fig. 1. Feynman diagrams for the lepton selfmass in the Weinberg-Salam-Gim model.
  (a) $B \equiv \gamma, Z^0, W^+, W^-$ (gauge fields); $s^+, s^-, \chi$ (unphysical scalars); $\psi$ (Higgs scalar)
  (b) $C \equiv W^+, W^-, Z^0, s^+, s^-, \chi, \psi; \phi^+, \phi^-, \phi^Z$ (ghosts); $f_j$ (charged leptons and coloured quarks).

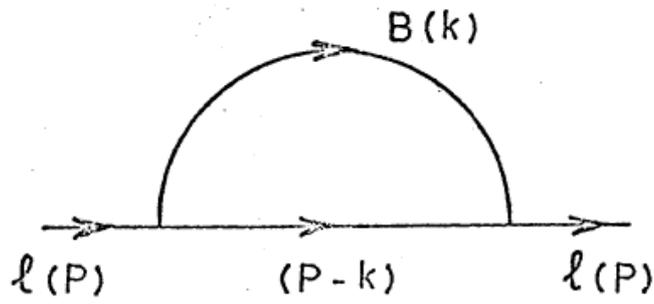

– fig. 1a –

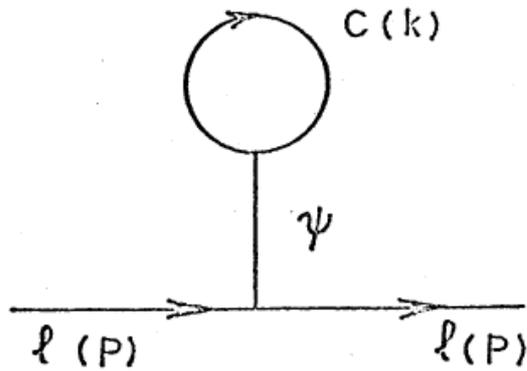

– fig. 1b –